\documentclass[aps,titlepage,12pt]{revtex4}
\usepackage{amsfonts}
\usepackage{amsmath}
\usepackage{graphicx}
\usepackage{dcolumn}
\usepackage{bm}
\usepackage{overpic}
\usepackage{booktabs}
\usepackage{color}
\usepackage[justification=raggedright,font={small,sf}, singlelinecheck=false]{caption}

\newcommand{\req}[1]{Eq.~(\ref{#1})}

\begin{document}
\title{Characteristics of human mobility patterns revealed by high-frequency cell-phone position data}

\author{Chen Zhao$^{1,2}$, An Zeng$^{3,}$\footnote{anzeng@bnu.edu.cn} and Chi Ho Yeung$^{4,}$\footnote{chyeung@eduhk.hk}}

\affiliation{
$^1$ College of Computer and Cyber Security, Hebei Normal University, Shijiazhuang, 050024, P. R. China\\
$^2$ Hebei Key Laboratory of Network and Information Security, Shijiazhuang, 050024, P. R. China\\
$^3$ School of Systems Science, Beijing Normal University, Beijing 100875, P. R. China\\
$^4$ Department of Science and Environmental Studies, The Education University of Hong Kong, Tai Po, Hong Kong
}

\begin{abstract}
Human mobility is an important characteristic of human behavior, but since tracking personalized position to high temporal and spatial resolution is difficult, most studies on human mobility patterns rely largely on mathematical models. Seminal models which assume frequently visited locations tend to be re-visited, reproduce a wide range of statistical features including collective mobility fluxes and numerous scaling laws. However, these models cannot be verified at a time-scale relevant to our daily travel patterns as most available data do not provide the necessary temporal resolution. In this work, we re-examined human mobility mechanisms via comprehensive cell-phone position data recorded at a high frequency up to every second. We found that the next location visited by users is not their most frequently visited ones in many cases. Instead, individuals exhibit origin-dependent, path-preferential patterns in their short time-scale mobility. These behaviors are prominent when the temporal resolution of the data is high, and are thus overlooked in most previous studies. Incorporating measured quantities from our high frequency data into conventional human mobility models shows contradictory statistical results. We finally revealed that the individual preferential transition mechanism characterized by the first-order Markov process can quantitatively reproduce the observed travel patterns at both individual and population levels at all relevant time-scales.
\end{abstract}


\maketitle

\section{Introduction}
Due to the increasing availability of mobile-phone records, global-positioning-system data and other datasets capturing traces of human movements, numerous statistical patterns in human mobility have been revealed, ranging from the confined radius of gyration at the individual level~\cite{understand2008Gonzalez} to the commuting fluxes at the collective level~\cite{modeling2010song}. These empirical observations suggest that human mobility are barely random, but follow predictable rules~\cite{scaling2006brockman,network2010eagle,predictability2011takaguchi,tword2011hu,universal2012simini,tale2012noulas,commuting2012Lenormand,seoul2012goh,persistence2013Saramaki,human2013Simini,
dynamic2014deville,memory2014hou,stochastic2016Gallotti}. Accordingly, models have been proposed to understand the observed mobility patterns. Following the pioneer model which generates empirical scaling behaviors by introducing two generic mechanisms, exploration and preferential return (EPR)~\cite{modeling2010song}, a large number of models for individual human mobility have been developed. Examples include the variants of the EPR model which describe user virtual mobility in cyberspace~\cite{understand2012szell,cyber2014zhao,unidied2016zhao} by incorporating a gravity model to simulate the returner-explorer dichotomy~\cite{returner2015pappalardo}, introducing a social circle to model the conserved number of locations an individual visits~\cite{evidence2018alessandretti}, aggregating individual trajectories to generate collective movements~\cite{universal2017yan}, and so on.

On the other hand, it has been shown that there is a diversity of human mobility patterns at different spatial scales. On the largest spatial scale which constitutes international movements, they are largely constrained by the entry requirement of individual countries, leading to asymmetric international movements~\cite{characterize2016li,predictability2012lu}. On the spatial scale within a country, models which describe international movements do not well explain inter-city movements. For instance, the inter-city human mobility is claimed to be mainly driven by the search for better job opportunities~\cite{universal2012simini,predicting2014ren}. The radiation model assumes that individuals tend to select the nearest locations with large benefits. On the spatial scale within a city, the local movements are better predicted by a population-weighted opportunity model where the potential area of coverage of individuals includes the whole city as a manifestation of the high mobility at the city scale~\cite{universal2014yan}. Although large efforts have been devoted to understand human mobility at different spatial scales, the studies of human mobility at different temporal scales are limited, due to the lack of high frequency mobility data~\cite{spatiotemporal2013hasan,partial2016geng}. Understanding spatial-temporal human mobility patterns at different scales would lead to numerous applications, such as suppressing epidemic spreading~\cite{natural2011belik,response2011bengtsson}, mitigating traffic congestion~\cite{timegeo2016jiang,addressing2018vazifeh}, urban planning~\cite{morphology2017lee,purpose2012alexander} and so on.

To reveal the human mobility pattern at different temporal scales, high frequency position data are required. While most existing empirical studies on human mobility are based on cell phone position data, these data are CDRs (Call Detail Records) where user positions are only recorded when they initiate or receive a call or a text message~\cite{survey2015blondel}. These datasets can include position records of up to several million anonymous mobile phone users, but the data has in general a low temporal resolution, as user positions are not recorded most of the time~\cite{burstiness2008goh}. The missing position data in some literature are interpolated via specific optimization algorithms or are incorporated from other data sources~\cite{travel2015toole,understand2016colak}. Difference may exist between the interpolated and the real data. Another usual practice to improve the temporal resolution of the data is to filter out users with long idle periods. For instance, this approach has been applied to extract a sample of user data with sufficient mobility records for inferring the nature of their visited locations such as home and workplace, and their tour trajectories with start and end point at home are investigated accordingly~\cite{timegeo2016jiang}. However, many problems still remain. On one hand, the user filtering procedure may lead to the risk of biased sampling of the original data. Specifically, the filtered data only include users who make frequent phone calls and may be biased to users with specific professions. On the other hand, the temporal resolution of the data after filtering is still insufficient (as frequent as every 10 min in existing literature), leaving many detailed user mobility traces missing from the data. Another possible data source is global-positioning-system (GPS) data~\cite{understand2016lima,diversity2013yan}. Their temporal resolution can be very high, but as GPS data are mostly recorded by navigation devices in vehicles, it only records positions when users are driving. As a result, GPS data are commonly used for analyzing traffic but not human mobility~\cite{speed2016wang}.

In this paper, we utilize the cell phone 4G communication data in a city in northern China to identify the location of individual cell phone user to a high frequency of every second. By incorporating various empirical quantities extracted from the data into the state-of-the-art human mobility models, we re-examine the models' capability in characterizing human mobility at different time-scales. A significant inconsistency between the real and the model behavior is found when the temporal resolution is high. We found that such inconsistency largely come from the assumption of a high tendency to re-visit locations that one has frequently visited, which holds in a long time-scale but not in a short time-scale. We further identify the critical time resolution above which existing models fail. Finally, we consider a simple model characterized by the first-order Markov process to quantitatively reproduce the observed travel patterns at both the individual and population levels in the high temporal resolution data. Our work reveals the heterogeneity in human mobility mechanism at different temporal scales, opening up a new dimension for understanding human mobility behaviours.

\section{Results}

\textbf{The high frequency human mobility dataset.} Our study is based on a full set of 4G communication data for 14 days between cell phones and cell towers in a city in northern China. The position of a user is recorded when his/her cell phone connects to the closest cell towers for the 4G communication service. As most applications in cell phones constantly exchange data with the back-end servers, the position of a user can be recorded up to every second. The original data include records of 5,336,194 users. In order to obtain a dataset describing the mobility patterns of active users with high temporal resolution, we have implemented strict rules to exclude users who do not move at all and those whose data is largely incomplete (i.e. those who have one or more days with less than 20-hour daily record in the consecutive 14-day period). Finally, we single out and analyze the mobility data of 55,389 users who satisfy the above criteria. The basic descriptive statistics of this data is shown in Fig.~S1 and S2 of the supplementary information (SI).

\textbf{Preferential return and empirical human mobility pattern at short time-scales.} We start our analysis by constructing the mobility network of a typical mobile phone user in Fig.~1a. Each node is a location visited and stayed more than 3 mins by the user, with its size proportional to the frequency he/she visited the location. Two nodes are connected by a link if the user has traveled at least once between the two locations. In the literature, there are numerous models which aim to reproduce the statistical properties of human mobility networks~\cite{modeling2010song,universal2012simini,predicting2014ren}. Two crucial mechanisms in the pioneer models are (i) exploration, i.e., the tendency of users to visit new locations, and (ii) preferential return (PR), i.e., the tendency of users to re-visit locations according to the frequency that they were visited in the past~\cite{modeling2010song}. As the tendency to explore new locations is set to decrease with time, the eventual dominant driving mechanism for human mobility in these models is preferential return, which successfully reproduces the power-law distribution of visitation frequency among locations and numerous other scaling laws observed in the empirical data. Many factors such as the population of the location and the aging effect are later introduced to obtain better agreement with empirical data~\cite{universal2017yan,evidence2018alessandretti}.

Here, we utilize our high frequency data to examine the limit of preferential return in explaining human mobility pattern. To this end, we reshuffle the trajectory of typical users by randomly reordering the sequence of their visited locations. The frequency users visited specific locations is therefore preserved. The mobility pattern constructed from the reshuffled trajectory of the typical user in Fig.~1a is illustrated in Fig.~1b. An obvious difference is observed when we compare Fig.~1a and 1b, suggesting that the preferential return mechanism adopted by most existing models fails to reproduce realistic mobility networks. Similar results of the real and the reshuffled trajectories of three other randomly selected users are shown in Fig.~S3 of the SI.

In order to quantify the statistical difference between the mobility patterns in real and reshuffled trajectories, we consider four metrics to quantify the trajectories of individuals.

The first one is the total number of unique transited location pairs (transited pairs for short), denoted as $n_\alpha^{\rm pair}$ for user $\alpha$,  which is equivalent to the number of links in the mobility network of user $\alpha$. We then compare $n_\alpha^{\rm pair}$ for all users in the real data and the reshuffled data in Fig.~1c. A box in the standard boxplots are marked in green if the line $y=x$ lies between 10\% and 91\% in each bin and in red otherwise. One can see that $n_\alpha^{\rm pair}$ in the reshuffled data is significantly larger than that in the real data. It is because for each individual there exist a few locations with large visitation frequency (e.g. home or office), in the reshuffled data users are attracted back to these locations regardless of the distance from the current location, before visiting other locations. In the real data, however, users do not always return to the frequently visited locations if they are too far away, resulting in a much smaller $n_\alpha^{\rm pair}$, i.e. a much fewer transited pairs than that in the reshuffled data.

The second metric we examined is the spread, as measured by the variance $Var_\alpha$, among the usage frequency of transited pairs of user $\alpha$. As shown in Fig.~1d, a large $Var_\alpha$ indicates that an individual $\alpha$ repeatedly uses a small number of routes and occasionally traveled through other routes. One can see that the values of $Var_\alpha$ are larger in the real data than in the reshuffled data, implying that users in the real data more frequently travel between a smaller number of location pairs.

The third metric we examined is the covered distance $d_\alpha^{\rm loop}$ of the maximum loop travelled by user $\alpha$. Here, a loop is defined as a trajectory that an individual starts from one location and ends in the same location. As shown in Fig.~1e, $d_\alpha^{\rm loop}$ is computed as the total geographic distance of the longest loop in each user's mobility trajectory. Larger $d_\alpha^{\rm loop}$ is observed in the reshuffled data, as users in the reshuffled data always return to the frequently visited locations even if they are far away.

Finally, the fourth metric, the total traveled distance $d_\alpha^{\rm total}$, is larger in the reshuffled data, as shown in Fig.~1f. As this metric is very sensitive to discrepancies in the predicted trajectory, it is largely ignored in the existing literature. The larger $d_\alpha^{\rm total}$ in the reshuffled data is also due to the fact that users often return to the far away yet frequently visited locations in the reshuffled data. In fact, $d_\alpha^{\rm total}$ is an important metric, capturing the geographic features of human mobility. All the above results suggest that although the reshuffled trajectories of individuals preserved the location visitation frequency, but if mobility patterns are merely explained by the preferential return mechanism, the patterns from the reshuffled data are significantly different from those in the real data. These results further imply that PR is not a good mechanism to describe individual mobility trajectories in a small time-scale.

Although PR does not explain well the mobility of individuals, one may wonder whether it is more valid at the collective level. To verify the validity of PR at the collective level, from each location, we compute the number of different locations that users travel to. This quantity is essentially the number of links that a location $i$ has in the mobility network, denoted as $k_i$. The corresponding distribution is shown in Fig.~1g. We see that both distributions $P(k_i)$ of the real and the reshuffled data resemble distributions with a power-law tail, yet their exponents are clearly different, with the tail obtained from the real data to be much shorter. We see similar difference when we compare the population flux $F_{ij}$ between each pair of locations $ij$ in the real data and the reshuffled data in Fig.~1h. Both distributions $P(F_{ij})$ roughly follow power-laws, with a larger exponent observed in the real data, indicating that the reshuffled data have underestimated the maximum flux between two locations.

Other than revealing human mobility patterns in the spatial dimension, our high frequency data also allow us to reveal the temporal dimension of human mobility activities. To this end, we denote the duration of each of a user's stay at a location as $t^{\rm stay}$, and examine the distribution $P(t^{\rm stay})$ over all users. As we can see in Fig. 2a, $P(t^{\rm stay})$ shows a power-law head and an exponential tail. The power-law head suggests that the duration of a stay at different locations is heterogeneous, and there are a large number of locations with relatively short duration of each stay. Note that these values of duration are sufficiently large, e.g. larger than 3 minutes (typical time for users to walk out of the several hundred meters radiation range of a cell tower), and are not pass-by locations. On the other hand, the exponential tail is mostly contributed by the duration when users stay or sleep at home.

As evident from Fig.~2a, many locations visited by users for a short time may have been neglected if the dataset do not have a high temporal resolution. Since our 4G cell phone data record user positions in every second, this allows us to examine data with different temporal resolution by data pruning. In order to examine how the mobility statistics are affected by the temporal resolution of the datasets, we consider a threshold and remove all the visited locations with $t^{\rm stay}<T$, for all users. In Fig.~2b, we show the average number of visited locations as a function of $T$. One can see that the number of visited locations decreases with an increasing $T$ in a power-law form, implying that the lower the temporal resolution of the data, the more substantial fraction of the visited locations are overlooked in the analyses. Indeed, many hidden mobility patterns at the short time-scale may have been neglected in existing studies which are based on mobility datasets with a low temporal resolution.

To further examine how the temporal resolution of the dataset affects the mobility statistics, we show in Fig.~2c-2f the difference between the real and the reshuffled data in terms of the total number of transited location pairs $n_\alpha^{\rm pair}$, the spread $Var_\alpha$ of the traveled frequency of transited pairs, the distance $d_\alpha^{\rm loop}$ covered by the maximum loop, the total traveled distance $d_\alpha^{\rm total}$, under various data removal thresholds $T$. The difference is measured by the fraction of users whose metric values in the reshuffled data are larger than those in the real data, except for $Var_\alpha$. As data reshuffling tends to decrease the spread of the traveled frequency of transited pairs, the difference in $Var_\alpha$ is computed as the fraction of user $\alpha$ with $Var_\alpha$ in the reshuffled data smaller than that in the real data. Remarkably, when temporal resolution is low (i.e. $T$ is large), our results only show a small difference between the real and the reshuffled data in terms of these four metrics at the individual level, in contrast to our findings in Fig.~1. This further suggests that the low temporal resolution of the dataset may have masked the discrepancy of the preferential return mechanism in describing the human mobility patterns at the short time-scale, leading to the seemingly good agreement between the preferential return mechanism and the observed human behaviors.

Similar results can be observed when we compare the power-law distributions in Figs.~1g and 1h under different temporal resolutions. Fig.~2g and 2h show that the difference between the exponents of the distributions in Fig.~1g and 1h obtained from the real and the reshuffled data is large when the threshold $T$ is small, then become negligible when $T$ is large. Another important observation in Fig.~2h is that the exponent magnitude of the flux distribution increases with $T$, indicating that the maximum flux between locations is higher in cases with large threshold. In other words, using datasets with a low temporal resolution would underestimate the flux between locations. Additionally, we study motifs in human travel trajectories~\cite{unravelling2013schneider} in Fig.~S4 and S5 (see discussion in SI note 3). A detailed comparison of the human travel motifs in the real data and reshuffled data shows that the reshuffling process does not significantly alter the motif distribution when $T$ is large, yet the difference between the motif distribution in the real data and the reshuffled data is substantial when $T$ is small.

\textbf{Origin-dependent preference on the next visiting location.} In order to understand the reasons underlying the observed difference between the real data and the reshuffled cases, we compare their matrices recording the travel frequency of a typical user between each location pair. The matrices are computed with the temporal resolution $T =$ 3 min, and are shown as heatmaps in Fig.~3a and 3b respectively for the real data and the reshuffled data. Some large values can be seen in the heatmap of the real data, which suggests that users tend to repeatedly transit between a small number of location pairs. However, this preference of transitions, or equivalently the preference of transited location pairs, cannot be captured in the reshuffled data.

We further examine the probability for the selected typical user to visit different locations starting from different origins in Fig.~3c. Different locations are indexed in the horizontal axis, with each blue curve corresponds to the probability to visit other locations from a specific origin; the black dashed line corresponds to the overall visitation probability distribution. Clearly, different blue curves peak at different locations, suggesting that the next location that a user visits is not always the most frequently visited ones, but instead strongly depends on his present location. Similarly, we show the visitation probability distribution for each starting location in the reshuffled data in Fig.~3d, of which the peaks of the blue curves are consistent with those of the black dashed lines. The comparison between Fig.~3c and 3d shows that in the real data, users' preference on the locations to be visited are dependent on their current location.

A more quantitative analysis can be made by computing the probability that the most frequently visited location $j^*_i $ from location $i$ is consistent with the overall most frequently visited location $j^*$, i.e. $p_{j^*_i=j^* }$. Fig.~3e shows the scatter plot and the bin average of $p_{ j^*_i=j^* }$ for each user in the real and the reshuffled data. Fig.~3f shows the distribution of $p_{ j^*_i=j^* }$ for all users in the real and the reshuffled data. Both figures show that $p_{ j^*_i=j^* }$ is smaller in the real data than that in the reshuffled data, again suggesting the origin-dependent preference on the locations to be visited.

\textbf{Data-integrated Models.} With the comprehensive cell-phone position dataset and based on our previous findings, we go on to examine the essential mechanisms underlying human mobility patterns. To achieve the goal, we plug various empirical quantities such as the popularity of locations and the frequency of transition between locations into existing human mobility models, and compare the emergent behavior from the models with empirical results.

We first start with the simplest \emph{preferential return} model of which the probability for an individual to visit a location is proportional to the frequency the location was visited in the past. In ref.~\cite{modeling2010song}, the exploration and preferential return (EPR) model was proposed. In this case, an individual has a probability $p\propto S^{-\gamma}$ to visit a new location, where $S$ is the number of visited locations, and a probability $1-p$ to return to a visited location with a probability proportional to its visited frequency. Here, since our studied dataset only includes locations within a city and most users only occasionally explore new locations within the short time period, i.e. two weeks, of the dataset,  we consider the case where $S$ is sufficiently large and the mobility of an individual is solely driven by the preferential return mechanism. We can thus write down the transition probability $p_{\alpha:i\rightarrow j}(t)$ of an individual $\alpha$ to travel from a location $i$ to a location $j$ at time $t$ to be
\begin{align}
\label{eq_ipr}
p_{\alpha:i\rightarrow j}^{\rm IPR}(t) \propto f_{\alpha:j}(t).
\end{align}
where $f_{\alpha:j}(t)$ is the empirical frequency that a location $j$ is visited by an individual $\alpha$ before time $t$. We call the above the \emph{individual preferential return} (IPR) mechanism. A simulated trajectory with \req{eq_ipr} to be the transition probability is shown in Fig.~4b, again compared with the real empirical trajectory shown in Fig.~4a. As we can see, many transitions absent in the empirical data are found in the simulated results. Furthermore, we consider a metric $d_\alpha^{\rm total}$ to examine statistically the validity of this model. We use $d_\alpha^{\rm total}$ because it is a geographic-aware metric which captures even small inaccurate predictions of paths in the users' travel trajectory. As shown in the scatter plots of $d_\alpha^{\rm total}$ in Fig.~4g, other than a specific individual, many of the simulated trajectories are longer than their counterparts in the empirical data, which may be a result of the transitions between more distant locations in simulations as in Fig.~4b. These results imply that the IPR mechanism is insufficient to explain human mobility patterns. Since the data of IPR are independent of origin, one may expect that origin-dependent transitions are indeed crucial in explaining mobility patterns.

While the preferential return model is over-simplified in explaining human movement, we then explore the significance of origin-dependent transitions in explaining mobility patterns. Since the individual frequency of transition between two locations is difficult to be modeled, many existing studies only utilize the average transition frequency over the population. Related models for predicting the average transition frequency over the population include the gravity model~\cite{gravity1990erlander}, radiation model~\cite{universal2012simini}, population-weighted opportunity model~\cite{universal2014yan} and so on. We call this the \emph{population preferential transition} (PPT) mechanism, of which the transition probability $p_{\alpha:i\rightarrow j}(t)$ is given by
\begin{align}
\label{eq_ppt}
p_{\alpha:i\rightarrow j}^{\rm PPT}(t) \propto f_{i\rightarrow j}(t),
\end{align}
where $f_{i\rightarrow j}(t)$ is the empirical frequency of which the population travel from location $i$ to $j$ before time $t$. As shown in Fig.~4c, the trajectory of this specific individual is dominated by paths which connect between near locations, reflecting the average behavior of the population to go to near and attractive locations~\cite{gravity1990erlander,universal2012simini,universal2014yan}. This trajectory in Fig.~4c is significantly different from the real trajectory in Fig.~4a. Consistently, we see in Fig.~4h that the simulation underestimates the real total travel distance $d_\alpha^{\rm total}$ for most individuals in the empirical data. These results imply that individuals travel to fulfill specific purposes by which short distance is not the main consideration. Although not surprising, the results suggest that the PPT mechanism is insufficient to explain the individual mobility patterns.

In a recent work \cite{universal2017yan}, a model combining the memory effect and the population-induced competition is proposed to simulate human mobility between locations based only on their population. Basically, individual mobility in this model is driven by both preferential return and collective mobility between locations. In order to test whether this model can generate realistic human mobility at high temporal resolution data, we consider a population-weighted individual preferential return model (PIPR) combining IPR and PPT, with the transition probability given by
\begin{align}
\label{eq_ipr_ppt}
p_{\alpha:i\rightarrow j}^{\rm PIPR}(t) \propto f_{\alpha:j}\times f_{i\rightarrow j}(t).
\end{align}
This model is actually a simplified version of the model proposed in ref.~\cite{universal2017yan}, where the collective mobility between locations as predicted by popularity distribution is replaced by the population preferential transition probability. As shown in Fig.~4d and 4i, although the trajectory and the total travel distance are more similar to the empirical data than merely IPR or PPT, they are still different from the real data as it substantially underestimates $d_\alpha^{\rm total}$ in the high temporal resolution human mobility data.

Inspired by the empirical observation in Fig.~3 that people tend to repeatedly transit between a small number of location pairs, we consider here another model based on the first-order Markov process that might explain the driving mechanism in the high temporal resolution human mobility. We call the mechanism the individual preferential transition (IPT). In this case, the transition probability $p_{\alpha:i\rightarrow j}(t)$ is given by
\begin{align}
\label{eq_ipt}
p_{\alpha:i\rightarrow j}^{\rm IPT}(t) \propto f_{\alpha:i\rightarrow j}(t),
\end{align}
where $f_{\alpha:i\rightarrow j}(t)$ is the empirical frequency of which individual $\alpha$ travels from location $i$ to $j$ before time $t$. As we can see in Fig.~4e, the simulated trajectory resembles the real trajectory shown in Fig.~4a. Other than this specific individual, we see in Fig.~4j that the simulated $d_\alpha^{\rm total}$ of each individual shows a more linear relation with their counterparts in the real data, compared to the above three models (see Fig. 4g, 4h and 4i respectively). These results may imply that the IPT mechanism is sufficient to explain the mobility patterns of individuals in high temporal resolution, leaving other factors of preferential return or population competition less essential.

A remarkable advantage of the state-of-the-art human mobility models is that they can reproduce collective human mobility by aggregating simulated individual mobility trajectories~\cite{universal2017yan}. One important metric that is usually used to examine this feature is the distribution $P(F_{ij})$ of the flux between locations. Fig.~4f presents respectively the fitted curves of the power-law flux distribution generated by IPR, PPT, PIPR and IPT models. We compare these fits with that of the real data (in high resolution, stay duration threshold $T=3$ mins) and the reshuffled data. As we can see, the exponent generated by the PIPR model is very close to that of the real data. However, the exponent generated by the IPT model is identical to that of the real data, suggesting that IPT can best reproduce the real flux distribution.

To understand more comprehensively the difference between the IPT and IPR models, we study several additional metrics, with the results summarized in SI note 4. At individual level, we examine three other metrics including the number $n_\alpha^{\rm pair}$ of transited location pairs, the variance $Var_\alpha$ of the transited pairs' usage frequency, and the distance $d_\alpha^{\rm loop}$ of maximum loop, as presented in Fig.~S6. While IPR can reproduce the number of transited location pairs similar to that in the real data, it underestimates $Var_\alpha$, and overestimates $d_\alpha^{\rm loop}$. In Fig.~S7, we study another metric at the collective level, namely the distribution $F(k_i)$ of the number of different locations that users travel to starting from location $i$. A longer tail generated by the IPR model indicates that IPR would overestimate the number of different locations that users travel to originated from a specific location. IPT outperforms IPR in reproducing these metrics at both individual and collective levels. We finally simulate respectively the IPR and the IPT models in a finite space of $M$ locations with no initial memory. The results confirm the advantage of IPT in reproducing the realistic human mobility patterns (see results in Fig.~S8 and the discussion in SI note 4). Taken together, the IPT model, integrated with quantities extracted from the comprehensive cell-phone position dataset, can well reproduce human mobility patterns with high temporal distribution that other models fail to capture.

\section{Discussion}
To summarize, we presented a comprehensive study of human mobility patterns in different temporal scales with a large sample of 4G cell phone data where the positions of users are recorded in each second. We surprisingly found that the existing models of preferential return agrees with empirical mobility patterns only at the long time-scale, but overestimates largely the total number of transited location pairs and the total traveled distance at short time-scale. The collective statistics such as the population flux between locations are also overestimated. This is due to the fact that these preferential return models assume the tendency for people to re-visit frequently visited locations regardless of the distance from those locations. Instead of preferential return, we found that in the high resolution human mobility data individuals exhibit clear preference on transitions between locations, which is determined by the frequency of the routes that have been used before. We finally study a simple model based on the first-order Markov process (called individual preferential transition) where the preference of users on paths are accumulated in a matrix and users move according to their preferred paths. The model can quantitatively reproduce the empirical travel patterns at both the individual and population levels up to the high temporal resolution of our empirical data.

Promising future directions include improving the model by introducing the decay of the preference on paths with time, which will result in a more realistic model where the frequently used paths of an individual evolve. In addition, one can empirically study the path preference matrix of individuals, which provides clues to various human mobility behaviors such as explorers and returners observed at the population level in the literatures~\cite{returner2015pappalardo}. Other directions include extending the present work to multiple spatial scales across cities or even countries~\cite{universal2012simini,universal2017yan}. The ultimate goal is to obtain a universal model that can be applied to explain the individual and collective human mobility patterns at different spatial and temporal scales. From the perspective of applications, one can study the overlap of users' preference in traveling paths in order to understand and suppress traffic congestion. Answering these questions would not only offer a better understanding of the fundamental mechanisms that underpin individual human mobility, but may also substantially improve our ability to predict and control collective traffic flux~\cite{urban2016pollock}.

\section{Materials and methods}
\textbf{Data.} The data analyzed in this paper is the full sample data of 4G communications among cell phones for 14 days in a city in northern China. As long as the 4G communication service is activated on the cell phones, the phones connect themselves to the closest cell towers and the position of users is recorded. As most applications in cell phones constantly exchange data with the back-end servers, the position of users can be recorded up to every second. Compared with the traditional cell phone data (CDRs) where the position of users is only recorded when they make phone calls, our obtained dataset is much higher in temporal resolution for analyzing individual mobility behavior. Due to the popularity of smart phones, our dataset actually covers a large proportion of population in the city. For privacy reasons, the data is anonymous and each user is assigned with a unique ID. The original data includes records of 5,336,194 users. We filter out the IDs which have only one position record (i.e. to remove those who do not move at all) and those who have one or more days with less than 20-hour daily record in the consecutive 14-day period (i.e. to have a dataset with an even higher quality). As a result, we obtain the high quality mobility data of 55,389 users.

\textbf{Model simulation.} We simulate the four models (i.e., IPR, PPT, PIPR, IPT) and compare their reproduced mobility patterns with those in the real data. Here, the initial configurations of these models are drawn from the real data. Specifically, $f_{\alpha:j}(t)$ in IPR, $f_{i\rightarrow j}(t)$ in PPT, $f_{\alpha:i\rightarrow j}(t)$ in IPT are set to be the values extracted from the empirical data. The vectors of $f_{\alpha: j}(t)$ for each user $\alpha$ in IPR and the matrices of $f_{\alpha:i \rightarrow j}(t)$ for each user $\alpha$ in IPT are then updated during the simulation. In the IPT model, $f_{\alpha:i \rightarrow j}(t)$ increases by $1$ if individual $\alpha$ travels from location $i$ to $j$ during the simulation. Similarly, in the IPR model, $f_{\alpha: j}(t)$ increases by $1$ if individual $\alpha$ visits location $j$ during the simulation. We stop the simulation for an individual $\alpha$ after he/she finishes the same number of travels as in his/her real data for 14 days.

We additionally simulate respectively the IPR and the IPT models in a finite space of $M$ locations, in which $N=6\times10^4$ individuals move $s$ steps ($M$ is a random number between $2$ and $350$, and $s$ is a random number between $50$ and $800$). All $f_{\alpha:i \rightarrow j}(t)$ in the IPT model and $f_{\alpha: j}(t)$ in the IPR model for individual $\alpha$ are set to be the same small value initially (i.e., $f_{\alpha:i \rightarrow j}(t)=1$ and $f_{\alpha: j}(t)=1$ for simplicity) and then updated during the process (see details in SI note 4). The results suggest that IPT outperforms IPR in reproducing the observed mobility patterns in the real data, see Fig.~S8.

\clearpage
\noindent  \textbf{Acknowledgments.} \\
This work is supported by the National Natural Science Foundation of China (Nos. 61703136, 61603046, and 61672206), the Natural Science Foundation of Beijing (No. L160008), the Natural Science Foundation of Heibei (No. F2017205064), and the Natural Science Foundation of Heibei Education Department (No. QN2017088). CHY acknowledges the Research Grants Council of Hong Kong Special Administrative Region, China (Project No. EdUHK 28300215, EdUHK 18304316 and EdUHK 18301217).

\noindent \\ \textbf{Author contributions.} \\
AZ and CZ designed the research, CZ performed the experiments, all authors analyzed the results and wrote the manuscript.\\

\noindent \textbf{Competing financial interests.} The authors declare no competing financial interests.\\

\noindent \textbf{Data and materials availability.} The raw data are not publicly available to preserve users' privacy under the Mobile Privacy Policy of China. Derived data supporting the findings of this study are available from the corresponding authors upon request. Due to the data security of participants, 14-days trajectory data cannot be shared freely, but are partly available to researchers who sign a confidentiality agreement and meet the criteria for access to confidential data.

\clearpage
\section*{Figures}
\begin{figure}[h!]
  \centering
  \includegraphics[width=16 cm]{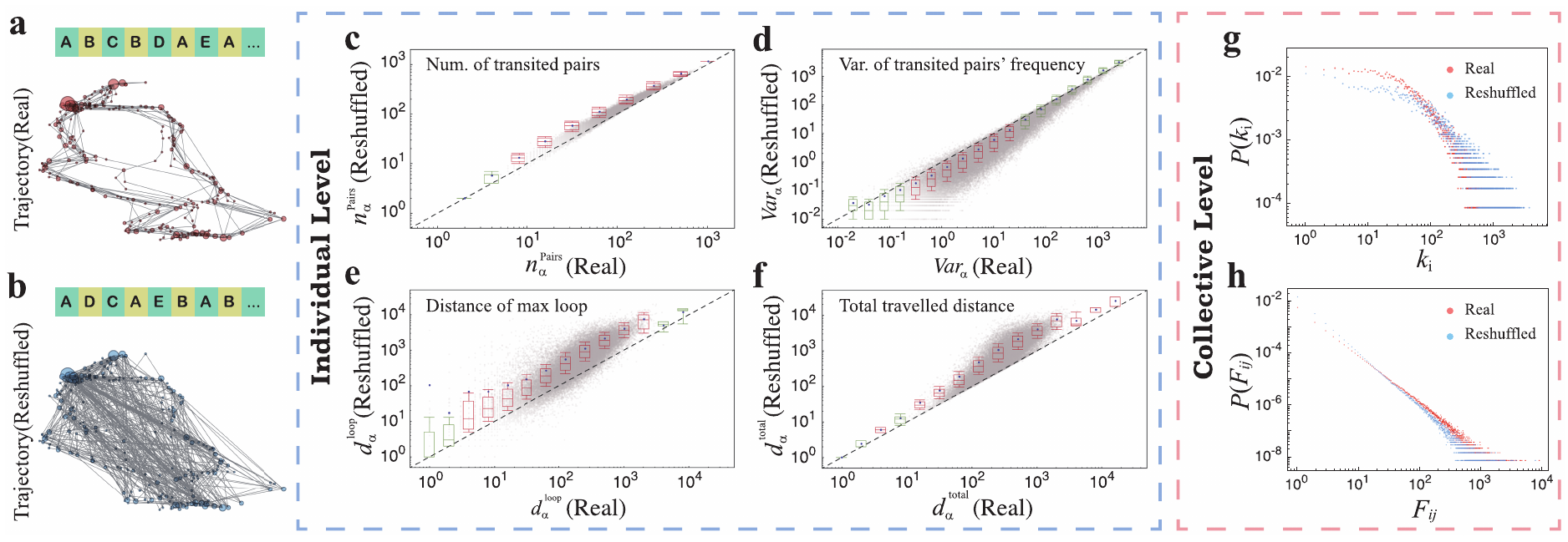}\\
  \caption{\textbf{Comparison of the statistical properties of the human mobility in  the real and the reshuffled data.} An illustration to compare (a) the real and (b) the reshuffled travel trajectory of a typical cell phone user. Here each node is a location visited by the user, with node size proportional to its visitation frequency. A link is drawn when the user has traveled at least once between the two locations. The reshuffled data is obtained by randomly reordering the sequence of visited locations. In this way, the visitation frequency of each location by the user is preserved while the travel trajectory is randomized. (c-f) Scatter plots comparing the statistical properties between the real data and the reshuffled data for each user $\alpha$ at the individual level, in terms of (c) the total transited location pairs $n_\alpha^{\rm pairs}$, (d) the variance $Var_\alpha$ of the traveled frequency of location pairs, (e) the covered distance $d_\alpha^{\rm loop}$ of the maximum loop, and (f) the total traveled distance $d_\alpha^{\rm total}$. A box in the standard boxplots are marked in green if the line $y=x$ lies between 10\% and 91\% in each bin and in red otherwise. (g-h) Comparison of the statistical properties of the real data and the reshuffled data at the collective level, in terms of the distributions of (g) the number of neighboring locations of each location, $P(k_i)$ (see (a) for example), and (h) the distribution of population flux between each two locations, $P(F_{ij})$.}\label{fig1}
\end{figure}

\begin{figure}[h!]
  \centering
  \includegraphics[width=16 cm]{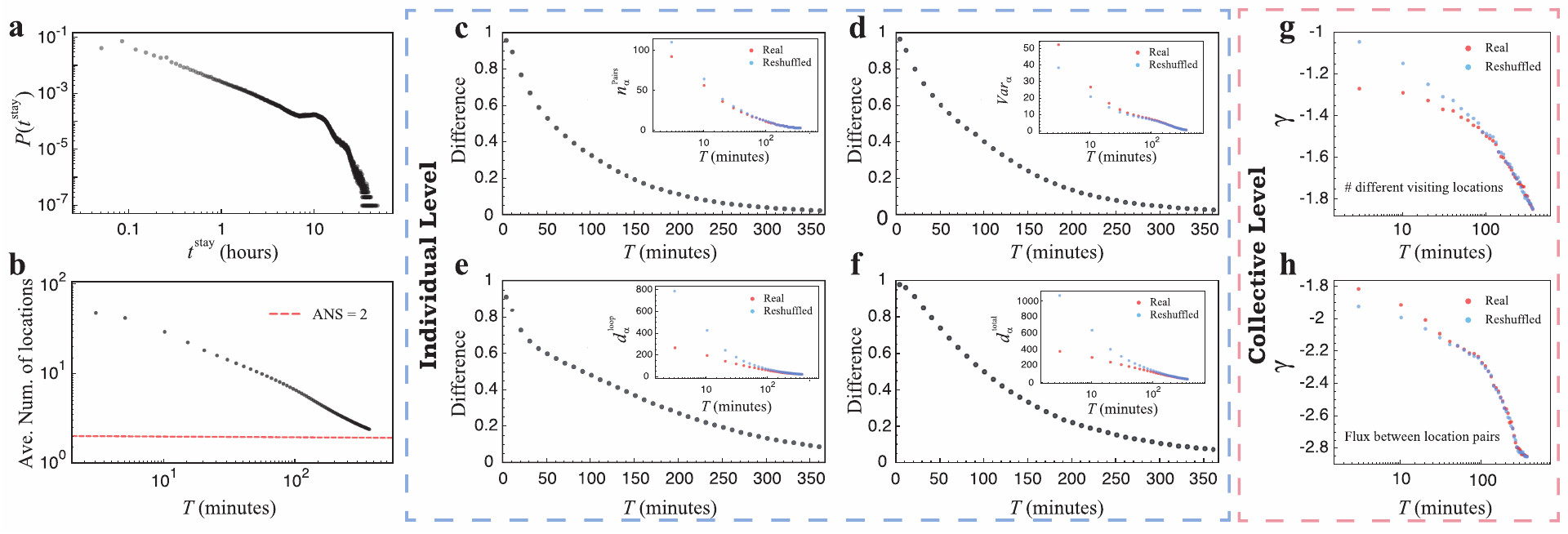}\\
  \caption{\textbf{The dependence of mobility statistical properties on the data temporal resolution.} (a) Distribution $P(t^{\rm stay})$ of the duration $t^{\rm stay}$ of a stay at each location for all users. We then conduct the statistics below by removing the visited locations with $t^{\rm stay}$ for all users. (b) The average number of visited locations as a function of $T$. In fact, the threshold $T$ can be regarded as a parameter controlling the resolution of the data. A larger threshold $T$ identifies less visited locations, and thus corresponds to a poorer data resolution. (c-f) The difference between the real data and the reshuffled data in terms of (c) the total transited location pairs $n_\alpha^{\rm pairs}$, (d) the variance $Var_\alpha$ of the traveled frequency of location pairs, (e) the covered distance $d_\alpha^{\rm loop}$ of the maximum loop, (f) the total traveled distance $d_\alpha^{\rm total}$, as a function of threshold $T$. The insets in each figure show the mean values of the corresponding metric under different threshold. The power-law exponents $\gamma$ of the distributions of (g) the number of neighboring locations from each location, and (h) population flux between each two locations, in the real data and the reshuffled data as a function of threshold $T$. In both (g) and (h), there is large difference between the exponents of the real and the reshuffled data when the threshold $T$ is small, and the difference becomes negligible when the threshold is large.}\label{fig2}
\end{figure}

\begin{figure}[h!]
  \centering
  \includegraphics[width=16 cm]{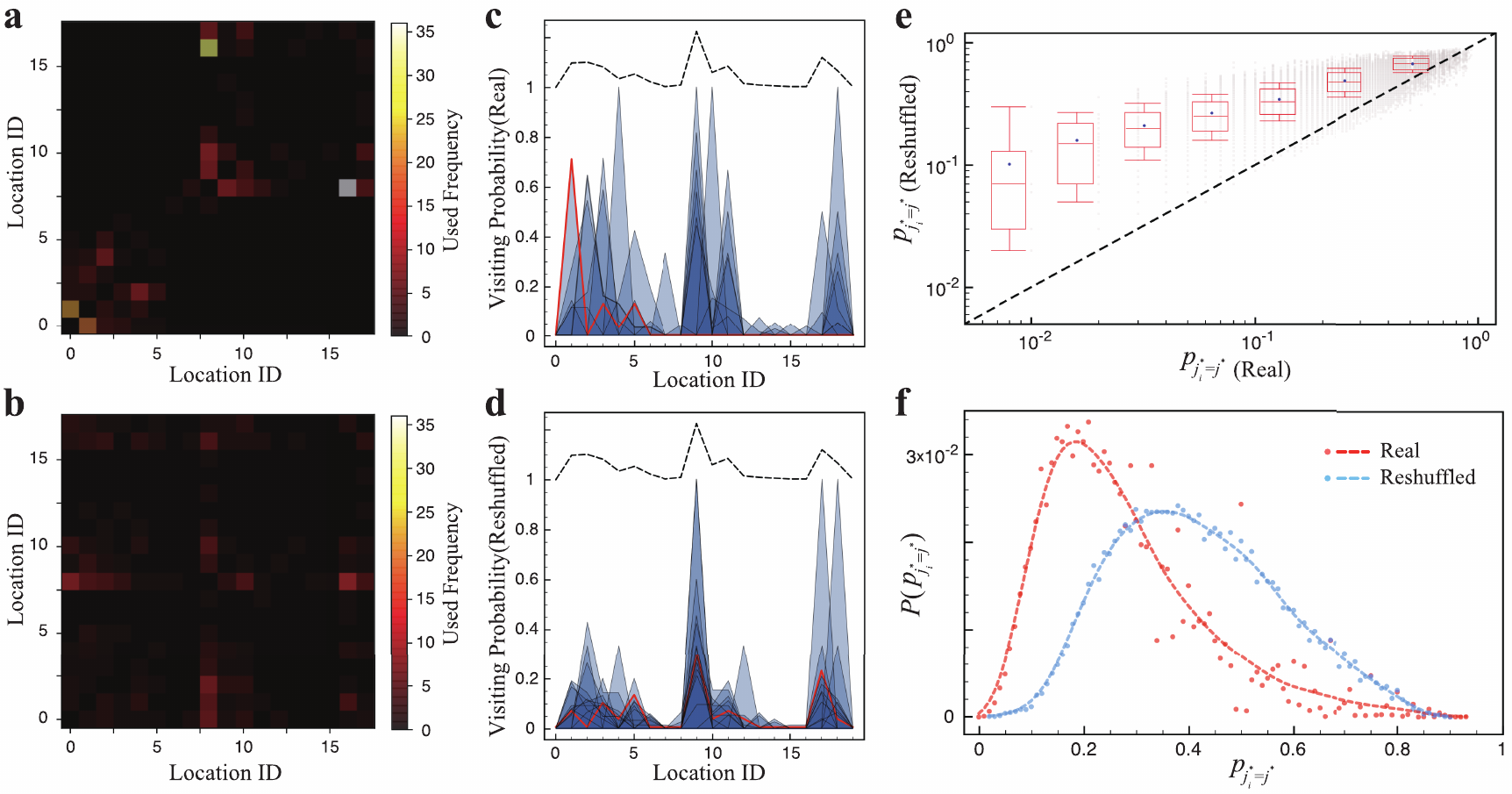}\\
  \caption{\textbf{Origin-dependent mobility behavior.} Heat maps which show the matrices of the travel frequency of a typical user from one location to another in (a) the real data, and (b) the reshuffled data of the selected typical user. The location visitation probability in (c) the real data and (d) the reshuffled data by the selected typical user originated from specific locations. As an example, the red curves in (c) and (d) shows the visitation probability distribution of the selected user originated from location 1, while the black curves show the visitation probability distribution aggregated from all starting locations. In (e) and (f), we show the probability of locations from which the most frequent locations to be next visited is the same as the overall most frequently visited locations (i.e. $p_{ j^*_i=j^*}$). The probability $p_{ j^*_i=j^*}$ is calculated for each user in both the real data and the reshuffled data. (e) shows the scatter plot of $p_{ j^*_i=j^*}$, indicating that in the real data the most likely locations to be next visited from many locations are different from the overall most frequently locations to be visited. (f) shows the distribution of $p_{ j^*_i=j^*}$ in real data and the reshuffled data.}\label{fig3}
\end{figure}

\begin{figure}[h!]
  \centering
  \includegraphics[width=16 cm]{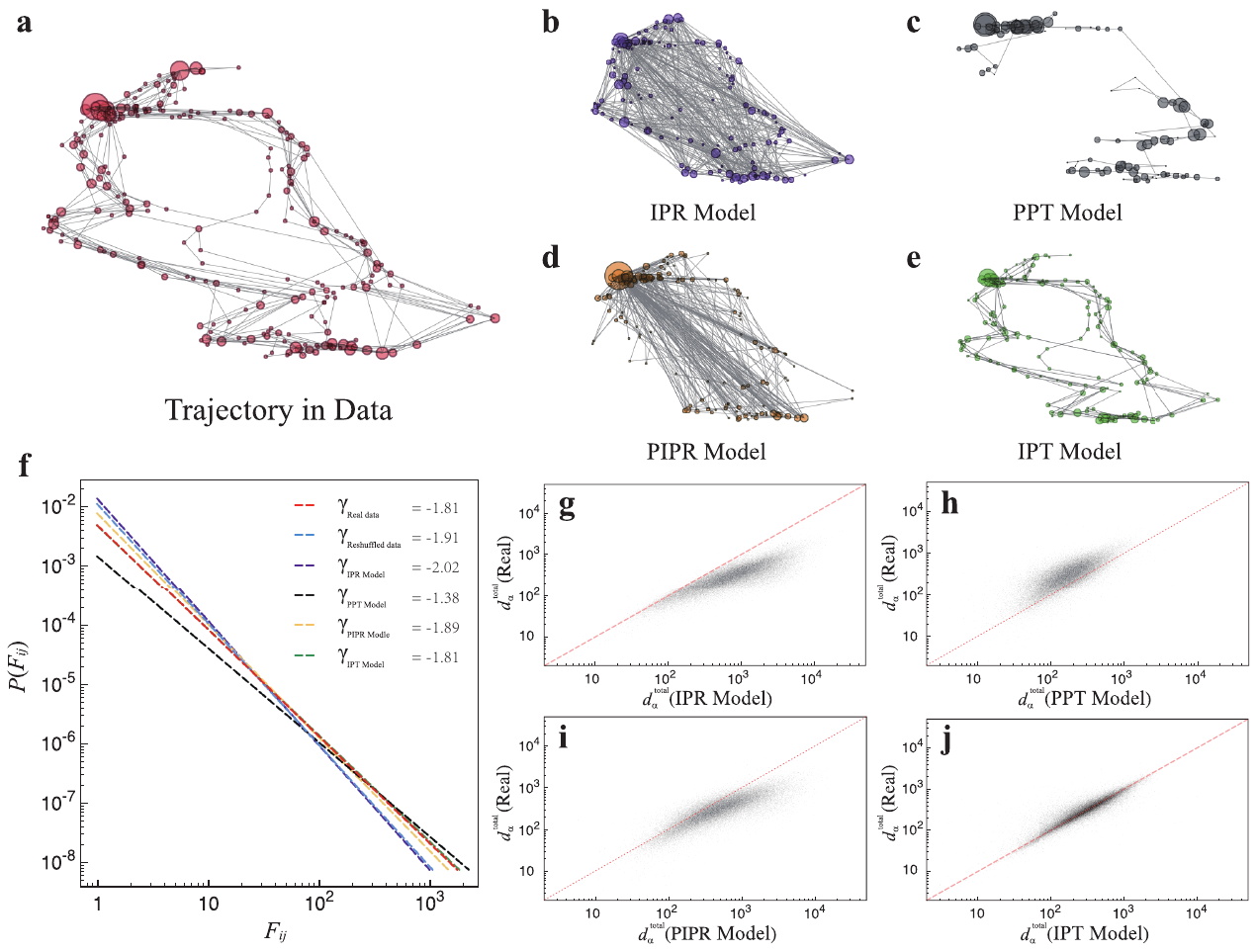}\\
  \caption{\textbf{Quantitative comparison of the IPT model with three representative models.} (a) The travel trajectory of a typical user in (a) the real data, and those simulated by (b) the IPR model, (c) the PPT model, (d) the PIPR model, and (e) the IPT model. For all the models, the relevant quantities are extracted from the real data of the selected user and used as the initial condition in the simulation. Here each node is a location visited by the user, with node size proportional to its visitation frequency generated by the corresponding model. (f) Comparison of the distributions of flux between location pairs, $P(F_{ij})$, in the real and the reshuffled data as well as the simulated trajectories from different models. The results of the fitted exponents suggest that the IPT model can best reproduce the flux distribution in real data. The scatter plots of the total traveled distance $d_\alpha^{\rm total}$ of each user $\alpha$ between the real data and the simulated data by (g) the IPR model, (h) the PPT model, (i) the PIPR model, and (j) the IPT model.}\label{fig4}
\end{figure}

\end{document}